\documentclass[11pt,a4paper,pdftex]{article}
\pdfoutput=1
\usepackage{jcappub}
\usepackage{color}

\newcommand{\be}{\begin{eqnarray}}
\newcommand{\ee}{\end{eqnarray}}

\title{Iron K$\alpha$ line of \\ Kerr black holes with scalar hair}

\author[a]{Yueying~Ni,}
\author[a]{Menglei~Zhou,}
\author[b]{Alejandro~C\'ardenas-Avenda\~no,}
\author[a,c,1]{Cosimo~Bambi,%
\note{Corresponding author}}
\author[d]{Carlos~A~R~Herdeiro,}
\author[d]{Eugen~Radu}

\affiliation[a]{Center for Field Theory and Particle Physics and Department of Physics,\\
Fudan University, 220 Handan Road, 200433 Shanghai, China}
\affiliation[b]{Programa de Matem\'atica, Fundaci\'on Universitaria Konrad Lorenz,\\ 
Carrera 9 Bis No. 62-43, 110231 Bogot\'a, Colombia}
\affiliation[c]{Theoretical Astrophysics, Eberhard-Karls Universit\"at T\"ubingen,\\ 
Auf der Morgenstelle 10, 72076 T\"ubingen, Germany}
\affiliation[d]{Departamento de F\'isica da Universidade de Aveiro and\\ 
Center for Research and Development in Mathematics and Applications (CIDMA),\\ 
Campus de Santiago, 3810-183 Aveiro, Portugal}

\emailAdd{yyni13@fudan.edu.cn}
\emailAdd{mlzhou13@fudan.edu.cn}
\emailAdd{alejandro.cardenasa@konradlorenz.edu.co}
\emailAdd{bambi@fudan.edu.cn}
\emailAdd{herdeiro@ua.pt}
\emailAdd{eugen.radu@ua.pt}

\abstract{Recently, a family of hairy black holes in 4-dimensional Einstein gravity minimally coupled to a complex, massive scalar field was discovered~\cite{hbh}. Besides the mass $M$ and spin angular momentum $J$, these objects are characterized by a Noether charge $Q$,  measuring the amount of scalar hair, which is not associated to a Gauss law and cannot be measured at spatial infinity. Introducing a dimensionless scalar hair parameter $q$, ranging from 0 to 1, we recover (a subset of) Kerr black holes for $q=0$ and a family of rotating boson stars for $q=1$. In the present paper, we explore the possibility of measuring $q$ for astrophysical black holes with current and future X-ray missions. We study the iron K$\alpha$ line expected in the reflection spectrum of such hairy black holes and we simulate observations with Suzaku and eXTP. As a proof of concept, we point out, by analyzing a sample of hairy black holes, that current observations can already constrain the scalar hair parameter $q$, because black holes with $q$ close to 1 would have iron lines definitively different from those we observe in the available data. We conclude that a detailed scanning of the full space of solutions, together with data from the future X-ray missions, like eXTP, will be able to put relevant constraints on the astrophysical realization of Kerr black holes with scalar hair.}


\begin{document}

\maketitle


\section{Introduction}

In 4-dimensional general relativity, the only stationary, regular on and outside an event horizon, asymptotically flat, vacuum black hole (BH) solution is the Kerr metric. This is the result of the so-called ``uniqueness theorems'', which were pioneered in Refs.~\cite{h1,h2,h3}; their final version is still an ongoing  research program (see, e.g., Ref.~\cite{hr}). This central result in BH physics, led to the presently accepted scenario that astrophysical BHs, which should form from the complete gravitational collapse of very massive stars, possess a spacetime geometry well described by the Kerr solution. Initial deviations from the Kerr metric are expected to be quickly radiated away by the emissions of gravitational waves immediately after the formation of the BH~\cite{price}. The equilibrium electric charge is negligible and cannot appreciably affect the background metric~\cite{ec}. The mass of the accretion disk is typically many orders of magnitude smaller than the mass of its BH and can be safely ignored~\cite{disk1,disk2}. Still, one crucial hypothesis that underlines this scenario, even within general relativity rather than extensions thereof, is that the Kerr solution the only ``physical" BH solution even when matter is consider, thus beyond the applicability of the vacuum uniqueness theorems. This is the content of the ``no-hair hypothesis"~\cite{Ruffini:1971bza}.

The currently available astrophysical data shows that BH candidates are dark and compact objects that can be naturally interpreted as the Kerr BHs of general relativity~\cite{ramnar}. Theoretical developments, however, question if they could be something else, in the presence of new physics, beyond vacuum general relativity. Indeed, either modifications of general relativity, or even simply the existence of new fundamental fields,  could lead to new BHs (or other compact objects), distinct from the paradigmatic Kerr solution. Consequently, in the past few years, there have been significant efforts to study how to test the Kerr BH hypothesis with electromagnetic radiation~\cite{p1,p2,p3,p4,p5,p6,p7,p8,p9,p10,p11,p12,p13,p14,p15,p16} and gravitational waves~\cite{gwt1,gwt2,gwt3,gwt4}; for a review, see, e.g., Refs.~\cite{r0,r1,r2,r3,gwt5}. Violations of the no-hair hypothesis are indeed possible in a number of cases. For instance, hairy BHs generically arise when gravity couples to non-Abelian gauge fields~\cite{vio1,vio2}, or when scalar fields non-minimally couple to gravity, e.g. a dilaton field in Einstein-dilaton-Gauss-Bonnet gravity~\cite{vio3}.

Recently, two of us have discovered a new family of asymptotically flat, hairy BHs in 4-dimensional Einstein gravity minimally coupled to a complex, massive scalar field~\cite{hbh}: \textit{Kerr BHs with scalar hair} (KBHsSH) -- see~\cite{Herdeiro:2015gia} for a detailed description,~\cite{Herdeiro:2014ima} for a contextualization, and~\cite{Kleihaus:2015iea,Herdeiro:2015tia,Herdeiro:2016tmi,Herdeiro:2014jaa,Cunha:2015yba} for generalizations and other physical properties. These solutions circumvent long standing no-scalar-hair theorems, i.e. no-go theorems for the existence of BHs in models beyond (electro)vacuum general relativity (see~\cite{Herdeiro:2015waa} for a review), by combining rotation with a harmonic time-dependence in the scalar field. The latter, however, cancels out at the level of the energy-momentum tensor; consequently these BHs have a stationary (and also axi-symmetric) geometry, just like the Kerr solution. In the limit when the scalar hair vanishes they reduce to a subset of Kerr BHs~\cite{Hod:2012px,hbh,Hod:2013zza,Hod:2014baa,Benone:2014ssa,Hod:2015ota}, whereas in the limit in which the horizon vanishes they reduce to well known horizonless gravitating solitons, known as boson stars~\cite{Schunck:2003kk,Liebling:2012fv}. The latter have been the subject of decades of research, and, in particular, have been suggested as dark matter candidates~\cite{Suarez:2013iw,Li:2013nal}. Being solutions of general relativity (rather than some alternative theory of gravity) minimally coupled to a simple matter model, obeying all energy conditions, and reducing to the paradigmatic general relativity BH in one limit and well known solitonic objects in another limit, KBHsSH provide a theoretically well motivated model for deviations from the Kerr paradigm. It is therefore interesting (and timely) to investigate how much the phenomenology of these hairy BHs can differ from that of Kerr.

In the present paper, we explore the possibility of constraining KBHsSH from the study of their X-ray reflection spectrum, with current and future X-ray missions. As a proof of concept, herein we only consider the iron K$\alpha$ line, which is the main feature in the reflection spectrum to study the spacetime geometry in the strong gravity field of the source, and we employ a small accretion disk. Moreover, we focus on a small sample of solutions, belonging to different regions in the parameter space of KBHsSH. This will be enough to show that significant differences from Kerr exist. A more detailed analysis, modeling the whole reflection spectrum and scanning the full space of solutions, will be left for future work.

The main challenge to compute the iron line in these spacetimes is that the metric is only known numerically. Here we use the code of Ref.~\cite{menglei}. The calculation time is significantly longer than the one necessary for an analytical metric. In order to study whether current and future X-ray mission can measure the scalar hair of these objects, we simulate the X-ray spectrum of the selected sample of hairy BHs and we fit the faked spectra with a Kerr model.

In our simulations, we consider Suzaku\footnote{http://heasarc.gsfc.nasa.gov/docs/suzaku/} and eXTP\footnote{http://www.isdc.unige.ch/extp/}, respectively as the prototype of a current and a future X-ray mission. When we obtain a good fit with a Kerr model, we conclude that we cannot distinguish the hairy BH under consideration from a Kerr BH without scalar hair. If the faked spectrum cannot be fitted with the Kerr model, however, we conclude that the scalar hair can be measured. Since current observations of the X-ray reflection spectrum of astrophysical BHs can be fitted with the standard Kerr models, we conclude that some KBHsSH are already at tension with current data. But we also observe that other regions of the parameter space, with significantly hairy BHs which could be observationally distinguished from the Kerr model using other observables -- such as their shadows~\cite{Cunha:2015yba,Vincent:2016sjq} --, are still allowed. We expect that future X-ray missions like eXTP will be able to put much stronger constraints on the scalar charge.

The content of the paper is as follows. In Section~\ref{s-2}, we briefly review the hairy BHs discovered in Ref.~\cite{hbh}. In Section~\ref{s-3}, we study the iron K$\alpha$ line expected in the reflection spectrum of hairy BHs. In Section~\ref{s-4}, we simulate some observations with Suzaku and eXTP, in order to figure out how present and future X-ray missions can test the possibility that astrophysical BHs have scalar hair. Summary and conclusions are presented in Section~\ref{s-5}. Throughout the paper, we employ natural units in which $c = G_{\rm N} =\hbar = 1$ and a metric signature $(-+++)$.

\section{Kerr black holes with scalar hair (KBHsSH) \label{s-2}}

KBHsSH are solutions to Einstein's gravity minimally coupled to a massive, complex, scalar field, with mass $\mu$. This model is summarized by the action:
 \begin{equation}
\label{actionscalar}
\mathcal{S}=\int  d^4x \sqrt{-g}\left[ \frac{R}{16\pi}
   -\frac{g^{\mu\nu}}{2} \left( \Phi_{, \, \mu}^* \Phi_{, \, \nu} + \Phi _
{, \, \nu}^* \Phi _{, \, \mu} \right) - \mu^2 \Phi^*\Phi
 \right]  \ .
\end{equation}
The BH solutions can be found using the following metric and scalar field ansatz:
\begin{equation}
 ds^2=-e^{2F_0}N dt^2+e^{2F_1}\left(\frac{dr^2}{N}+r^2 d\theta^2\right) + e^{2F_2}r^2 \sin^2\theta \left(d\varphi-W dt\right)^2 \ ,
 \label{kerrnc}
\ \ \ \ 
N\equiv 1 -\frac{r_H}{r} \ ,
\end{equation} 
\begin{equation}
\Phi(t,r,\theta,\phi)=e^{-iwt}e^{im\varphi}\phi(r,\theta) \ .
\label{sfansatz}
\end{equation}
Here, $r_H>0$ is the location of the event horizon, $w\in \mathbb{R}^+$ is the scalar field frequency and $m\in \mathbb{Z}^+$ is the azimuthal harmonic index. There is actually an infinite countable number of families of KBHsSH, each family being obtained for different values of $m$ and also of the number of nodes, $n$, of the scalar field profile $\phi$, along the equatorial plane. Here we shall focus on the nodeless family, $n=0$, and with $m=1$. These are likely the most stable configuration, as increasing either of these integer numbers (roughly) increases the energy of the solutions, corresponding to more excited states.

Even though the scalar field is complex, its energy-momentum tensor is real, and so is the geometry. This  geometry is described by four functions,  $F_i,W$, of the spheroidal coordinates $(r,\theta)$, whereas the scalar field introduces a fifth function of these same coordinates. These functions are found, numerically, by solving a system of five non-linear coupled partial differential equations, with appropriate boundary conditions that guarantee asymptotic flatness and regularity on and outside the horizon (see~\cite{Herdeiro:2015gia} for all the details). Of crucial importance is the \textit{synchronization condition}, $w=m\Omega_H$, where $\Omega_H$ is the angular velocity of the horizon, computed from the above metric as the horizon value of the function $W$. This condition guarantees regularity of a non-trivial scalar field at the horizon, and has a very clear physical interpretation in the context of the superradiance phenomenon of Kerr BHs~\cite{hbh}. We remark that the vacuum Kerr metric can be written in this coordinate system, which differs from the standard Boyer-Lindquist coordinates by a radial shift. The explicit form of the coefficients of the Kerr metric in this coordinates can be found in~\cite{Herdeiro:2015gia,Herdeiro:2016tmi}. 

\begin{figure}
\begin{center}
\includegraphics[type=pdf,ext=.pdf,read=.pdf,width=11cm]{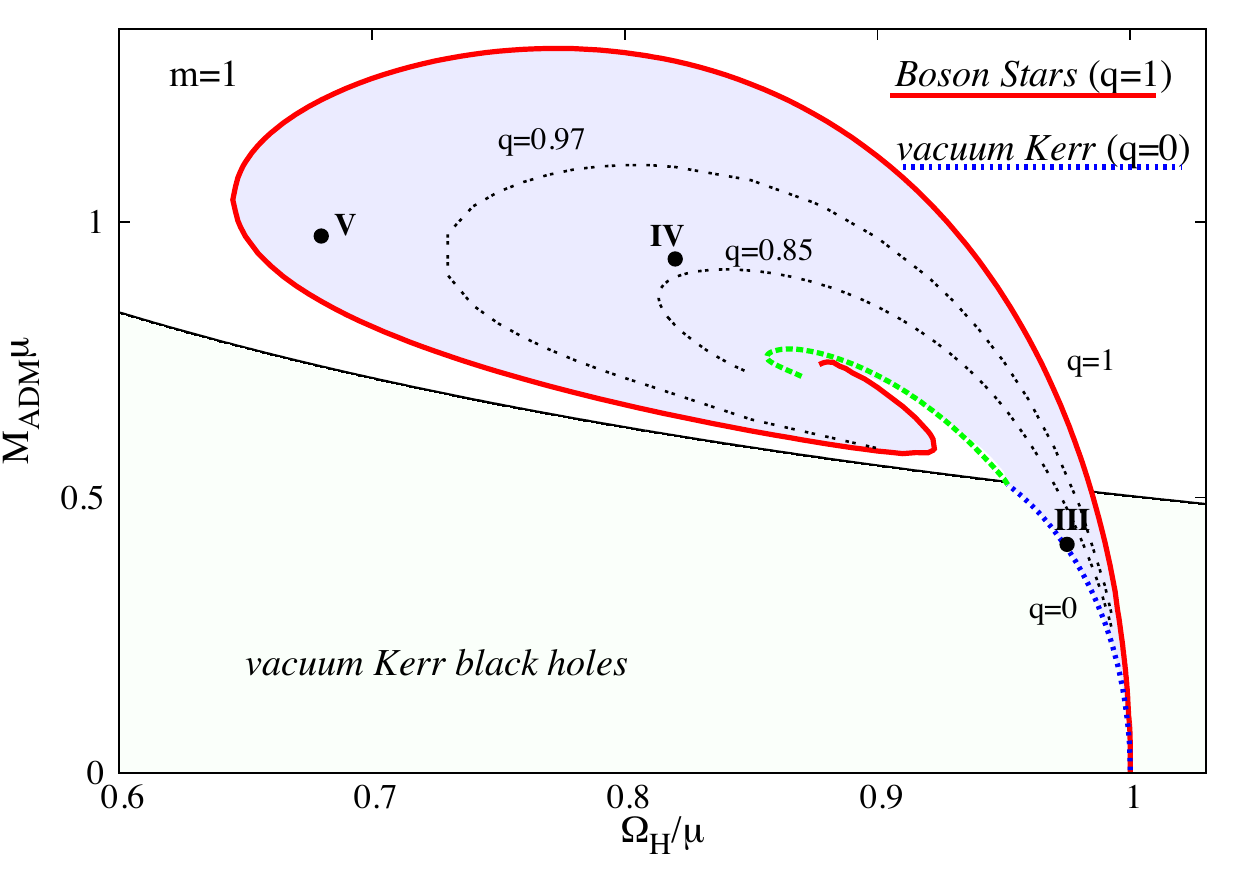}
\end{center}
\vspace{-0.5cm}
\caption{Domain of existence of KBHsSH (blue shaded region) in an ADM mass (here denoted $M_{\rm ADM}$ to avoid ambiguities) $vs.$ horizon angular velocity diagram. Vacuum Kerr BHs exist below the black solid line that corresponds to the extremal Kerr solution. KBHsSH bifurcate from the vacuum Kerr solution at a particular existence line, corresponding to the Kerr BHs that can support stationary bound states of the massive Klein Gordon equation (blue dotted line)~\cite{hbh}. This is the $q=0$ limit. In the $q=1$ limit, KBHsSH reduce to boson stars (red solid line). Constant $q$ lines are roughly parallel to the boson star spiral. The green dashed line is the set of extremal (zero temperature) KBHsSH.}
\label{domain}
\end{figure}

A generic KBHSH is described by three ``charges". The first two can be separated in their horizon (BH) and scalar field (S) contribution. They are the ADM or total mass $M=M_{\rm BH}+M_{\rm S}$, and the total angular momentum $J=J_{\rm BH}+J_{\rm S}$.\footnote{By total angular momentum we mean the angular momentum computed as the Komar integral associated to axi-symmetry, at spatial infinity.} The third one is a Noether charge, $Q$, which is conserved in a local sense (of a continuity equation), but, unlike the mass and angular momentum,  has no associated Gauss law, and hence cannot be measured by an observer at infinity. This Noether charge is a consequence of the $U(1)$ global symmetry of the complex scalar field and provides a measure of the hairiness of the BH. A given family of KBHsSH ($i.e.$ with a specific value of $m$ and $n$) bifurcates from a subset of vacuum Kerr BHs -- Fig.~\ref{domain}. In this limit, $M_{\rm S}=0=J_{\rm S}$ and $Q=0$. When the horizon size vanishes, which in particular implies $r_H\rightarrow 0$, the hairy BHs reduce to rotating boson stars (with the same values of $m$ and $n$, as boson stars also form an infinite countable number of families). In this limit, $M_{\rm BH}=0=J_{\rm BH}$ and $Q=mJ_{\rm S}$. This latter condition, which was first observed for rotating boson stars in~\cite{Schunck:1996he,Yoshida:1997qf}, also holds for KBHsSH. Thus we can write, for the hairy BHs,
\begin{equation}
1=\frac{J_{\rm BH}}{J}+\frac{Q}{mJ}  \ \ \Leftrightarrow \ \ \  q=1-\frac{J_{\rm BH}}{J} \ , \qquad q\equiv \frac{Q}{mJ}
\end{equation}
where we defined the normalized Noether charge $q\in [0,1]$, which measures the fraction  of angular momentum in the scalar field. This parameter provides a useful and compact measure of the hairiness, with $q=0$ for vacuum Kerr BHs (no hair) and $q=1$ for boson stars (only hair).

Since the hairy BH solutions are only known numerically, here we shall focus our attention on three specific KBHsSH, that were already studied for other phenomenological features, namely their shadows~\cite{Cunha:2015yba,Vincent:2016sjq}. The data for these solutions are publicly available\footnote{The data files are available at http://gravitation.web.ua.pt/index.php?q=node/416}; they are labeled and described as:
\begin{enumerate}
\item Configuration~III $(q=0.128)$ is a Kerr-like hairy BH. Only a relatively small fraction of the total mass and angular momentum is stored in the scalar field ($5\%$ of the mass and $13\%$ of angular momentum). The input parameters used in obtaining this solution are: $r_{\rm H} = 0.2$ and $\Omega_{\rm H} = 0.975$. In units of the scalar field mass\footnote{That is we have rescaled, in this discussion, $\mu r_H\rightarrow r_H$, $\Omega_H/\mu\rightarrow \Omega_H$, $\mu M\rightarrow M$ and $\mu^2 J\rightarrow J$. Observe that any solution can have any physical mass for an appropriate choice of $\mu$.}, the ADM mass is $M = 0.415$ and the total angular momentum is $J = 0.172$. The BH mass is $M_{\rm BH} = 0.393$ and the BH angular momentum is $J_{\rm BH} = 0.15$. The scalar field mass is $M_{\rm S} = 0.022$ and the scalar field angular momentum is $J_{\rm S} = 0.022$.
\item Configuration~IV $(q=0.846)$ already departs considerably from the vacuum Kerr solution. The largest portion of its mass and angular momentum are in the scalar field ($75\%$ of the mass and $85\%$ of the angular momentum). The input parameters are: $r_{\rm H} = 0.1$ and $\Omega_{\rm H} = 0.82$. The ADM mass is $M = 0.933$ and the total angular momentum is $J = 0.739$. The BH mass is $M_{\rm BH} = 0.234$ and the BH angular momentum is $J_{\rm BH} = 0.114$. The scalar field mass is $M_{\rm S} = 0.699$ and the scalar field angular momentum is $J_{\rm S} = 0.625$.
\item Configuration~V $(q=0.998)$ is an extreme departure from the vacuum Kerr solution. It describes, essentially a rotating boson star, wherein a tiny horizon exists. The mass and angular momentum are almost  fully in the scalar field ($98.2\%$ of the mass  and  $97.6\%$ of the angular momentum). The input parameters are: $r_{\rm H} = 0.04$ and $\Omega_{\rm H} = 0.68$. The ADM mass is $M = 0.975$ and the total angular momentum is $J = 0.85$. The BH mass is $M_{\rm BH} = 0.018$ and the BH angular momentum is $J_{\rm BH} = 0.002$. The scalar field mass is $M_{\rm S} = 0.957$ and the scalar field angular momentum is $J_{\rm S} = 0.848$.
\end{enumerate}

A representation of the location of these three solutions in the domain of existence of KBHsSH can be found in Fig.~\ref{domain}.

\section{Iron K$\alpha$ line of KBHsSH \label{s-3}}

Within the disk-corona model~\cite{corona1,corona2}, a BH is surrounded by an optically thick and geometrically thin accretion disk. The disk radiates like a blackbody locally, and as a multi-color blackbody when integrated radially. The temperature of the disk depends on the BH mass and mass accretion rate, and for thin disks it is in the X-ray band for stellar-mass BHs and in the UV/optical band for the supermassive ones. The so-called ``corona'' is a hotter ($\sim 100$~keV), usually optically thin, electron cloud, which enshrouds the central disk. From the inverse Compton scattering of the thermal photons from the disk off the hot electrons in the corona, the latter becomes an X-ray source with a power-law spectrum. The corona illuminates also the accretion disk, producing a reflection component with some emission lines, the most prominent of which is usually the iron K$\alpha$ line. The energy of this line is at 6.4~keV in the case of neutral iron, and it shifts up to 6.97~keV for fully ionized iron.

The iron K$\alpha$ line is very narrow in energy, while the one observed in the X-ray reflection spectrum of BH candidates is broad and skewed, as the result of relativistic effects (gravitational redshift, Doppler boosting, light bending) in the strong gravity region near the compact object~\cite{i1,i2,i3}. In the presence of the correct astrophysical model and high quality data, the study of the iron line profile can be a powerful tool to probe the spacetime metric around the BH candidate~\cite{i4,i5,i6}. In practice, one has to fit the whole reflection spectrum (as well as the other components of the spectrum). However, the iron K$\alpha$ line is usually the strongest feature to study the strong gravity region. For this reason, this technique is usually called iron line method.

The photon flux measured by a distant observer (in, e.g., units of s$^{-1}$~cm$^{-2}$~keV$^{-1}$) can be written as
\be
N_{E_{\rm obs}} = \frac{1}{E_{\rm obs}} 
\int I_{\rm obs} (E_{\rm obs}) d\Omega_{\rm obs}
= \frac{1}{E_{\rm obs}} 
\int g^3 I_{\rm e} (E_{\rm e}) d\Omega_{\rm obs} \, .
\ee
Here $I_{\rm obs}$ and $E_{\rm obs}$ are, respectively, the specific intensity of the radiation and the photon energy measured by the distant observer. $d\Omega_{\rm obs}$ is the line element of the solid angle subtended by the image of the disk on the observer's sky. $I_{\rm e}$ and $E_{\rm e}$ are, respectively, the local specific intensity of the radiation and the photon energy in the rest frame of the gas in the disk. $g = E_{\rm obs}/E_{\rm e}$ is the redshift factor and $I_{\rm obs} = g^3 I_{\rm e}$ follows from the Liouville's theorem.

The calculations of the iron K$\alpha$ line can be summarized as follows~\cite{p5,p6}. We consider a grid on the image plane of the distant observer and we fire photons from every point in the grid. We integrate backward in time the photon trajectories from the point of detection in the image plane of the distant observer to the point of emission in the disk. We employ the standard Novikov-Thorne model~\cite{nt}. The accretion disk is on the equatorial plane perpendicular to the BH spin. The particles of the gas follow nearly geodesic equatorial circular orbits. The inner edge of the disk is at the innermost stable circular orbit (ISCO) and we neglect any emission of radiation inside the ISCO. This is motivated by the fact that, once a particle reaches the ISCO, it quickly plunges onto the central BH.

With the ray-tracing, we can associate an emission point in the accretion disk to a point on the image plane of the distant observer and reconstruct the apparent image of the disk. Any point of the apparent image of the disk is characterized by its redshift factor $g$, which is determined by the point of emission and the photon initial conditions:
\be
g = \frac{\sqrt{- g_{tt} - 2 \Omega g_{t\phi} - \Omega^2 g_{\phi\phi}}}{1 + \lambda \Omega} \, .
\ee
Here $g_{\mu\nu}$ are the metric coefficient of the background metric, $\Omega$ is the Keplerian angular momentum of the gas in the disk (for a given metric, it only depends on the radial coordinate), and $\lambda = k_\phi/k_t$ is a constant of motion along the photon path, where $k_\phi$ and $k_t$ are the components of the photon 4-momentum. Integrating over the apparent image of the disk, we can get the observed spectrum of the source.

The iron line profile is determined by the background metric (in the Kerr metric, only by the spin parameter $a_* = J/M^2$), the inclination angle of the disk with respect to the line of sight of the distant observer $i$, and the local spectrum $I_{\rm e}$. The latter is often modeled with a power-law profile, say $I_{\rm e} \propto 1/r^\alpha$, where $\alpha$ is the emissivity index, or by a broken power-law, in which $I_{\rm e} \propto 1/r^{\alpha_1}$ for $r < r_{\rm b}$ and $I_{\rm e} \propto 1/r^{\alpha_2}$ for $r > r_{\rm b}$. In the second case, we have two emissivity indexes, $\alpha_1$ and $\alpha_2$, and a breaking radius, $r_{\rm b}$, to be inferred by fitting the data.

\begin{figure}
\begin{center}
\includegraphics[type=pdf,ext=.pdf,read=.pdf,width=11cm]{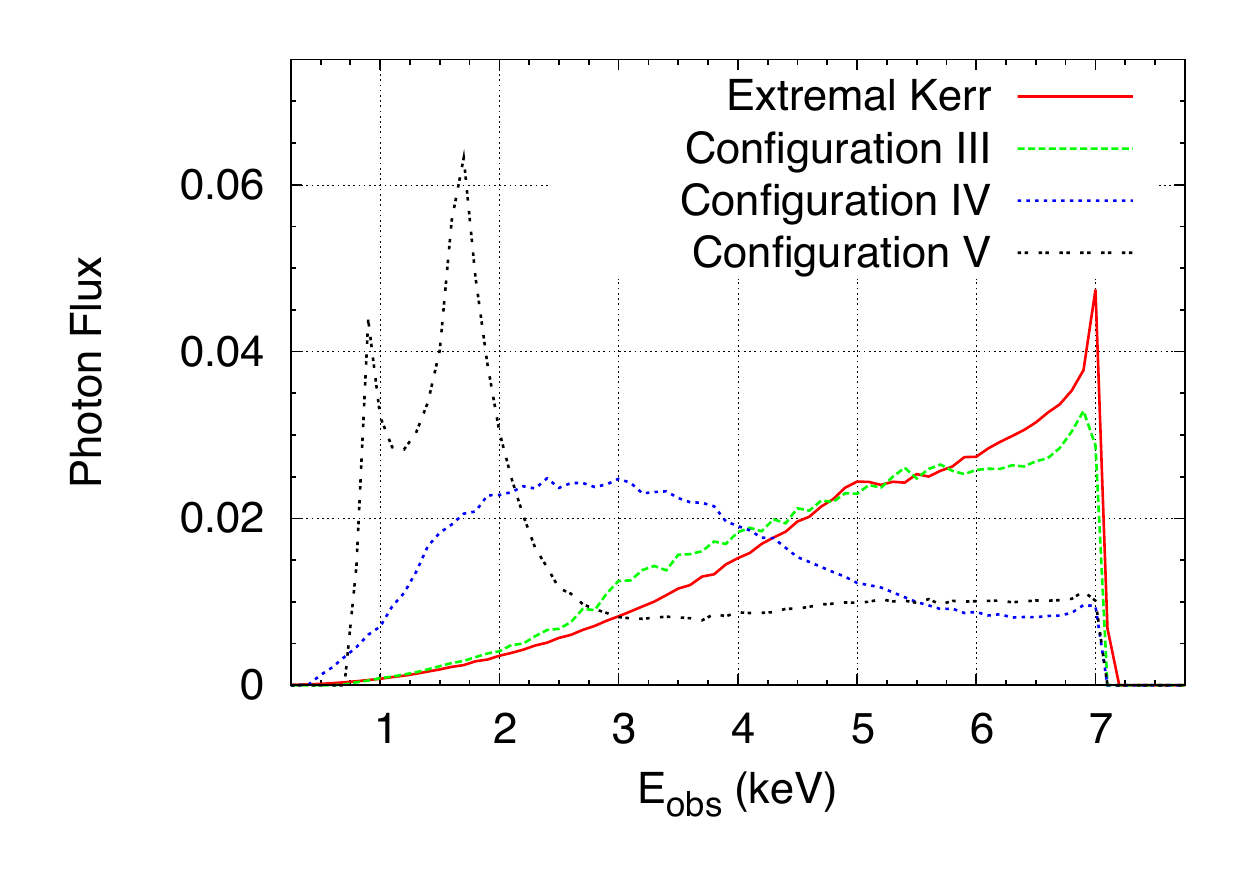}
\end{center}
\vspace{-0.8cm}
\caption{Iron line profiles of an extremal Kerr BH and of the three KBHsSH discussed in the present paper. The ``wiggles'' appearing in the shape of these lines, in particular for the configurations~III and IV, are due to resolution effects of the numerical metric. \label{firon}}
\vspace{0.3cm}
\begin{center}
\includegraphics[type=pdf,ext=.pdf,read=.pdf,width=11cm]{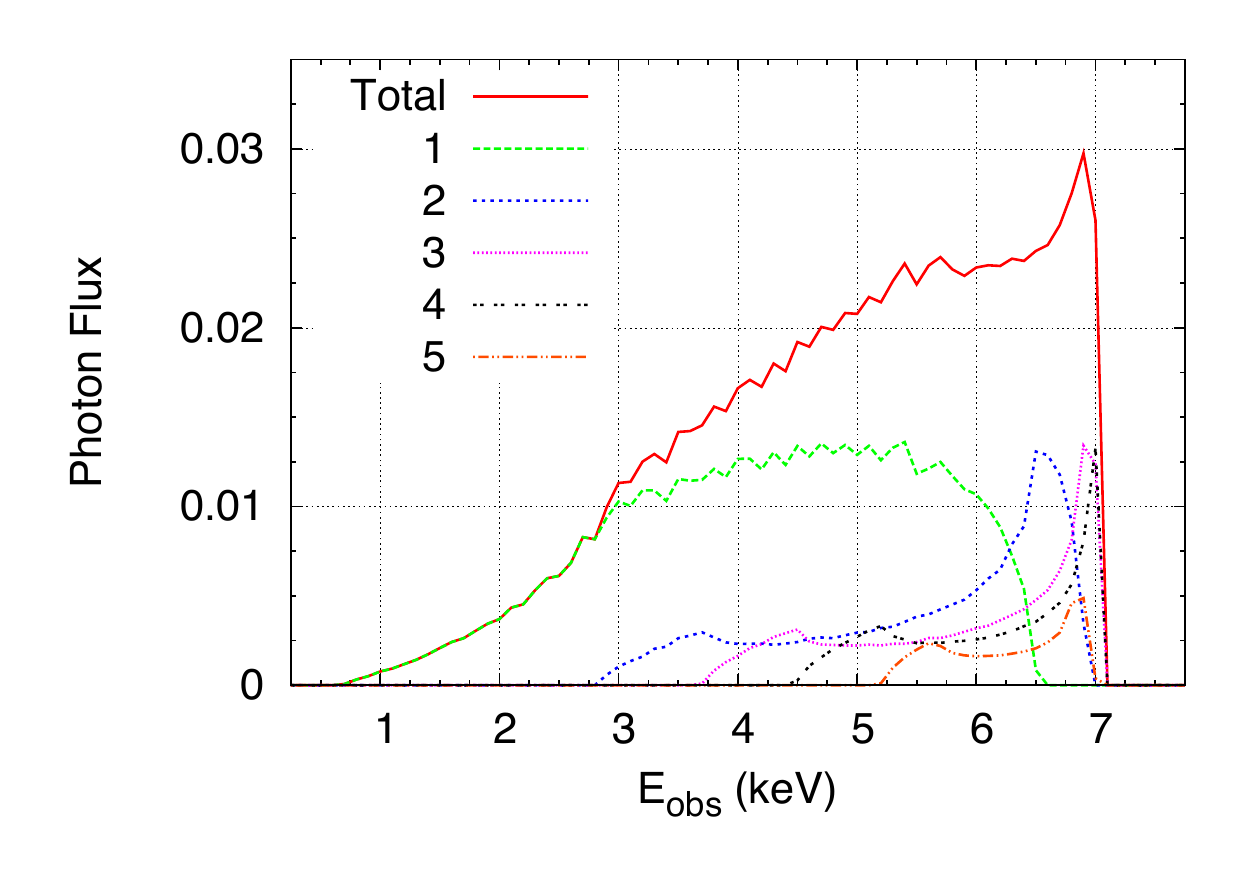}
\end{center}
\vspace{-0.8cm}
\caption{Iron line profile from the whole disk (red solid line) and contributions to the total iron line profile from different annuli (lines 1, 2, 3, 4, and 5) for the configuration~III KBHSH. The annulus~1 has the inner edge at the ISCO and the outer edge at the radius $r = r_{\rm ISCO} + 1$. The annulus~2 has the inner edge at the same radius as the outer edge of the annulus~1 and the outer edge at the radius $r = r_{\rm ISCO} + 2$. The annulus~3 has the inner edge at the same radius as the outer edge of the annulus~2 and the outer edge at the radius $r = r_{\rm ISCO} + 4$. The annulus~4 has the inner edge at the same radius as the outer edge of the annulus~3 and the outer edge at the radius $r = r_{\rm ISCO} + 10$. The annulus~5 has the inner edge at the same radius as the outer edge of the annulus~4 and the outer edge at the radius $r = r_{\rm ISCO} + 25$, which corresponds to the outer edge of the disk. \label{fc3}}
\end{figure}

The calculations of the iron line in the metrics described in the previous section have been done with the code described in Refs.~\cite{p5,p6} and extended in Ref.~\cite{menglei} for numerical metrics. The resulting iron line profiles are shown in Fig.~\ref{firon}, where we have employed a viewing angle $i=45^\circ$ and a simple power law with $\alpha = 3$ for the intensity profile. The latter corresponds to the Newtonian limit in the case of lamppost geometry~\cite{dauser}. 
The viewing angle $i$ mainly affects the Doppler boosting, so it changes the observed redshift $g$. The intensity profile $I_{\rm e}$ determines the relative contribution in the total line from every radius, and therefore it affects the shape of the line but not the photon redshift. Both parameters are important in a fit. We do not expect, however, that our choice can change the conclusions of this paper, because the aim of this work to figure out qualitatively different features that cannot be mimicked by a simple adjustment of the values of the parameters of the model.

\begin{figure}
\begin{center}
\includegraphics[type=pdf,ext=.pdf,read=.pdf,width=11cm]{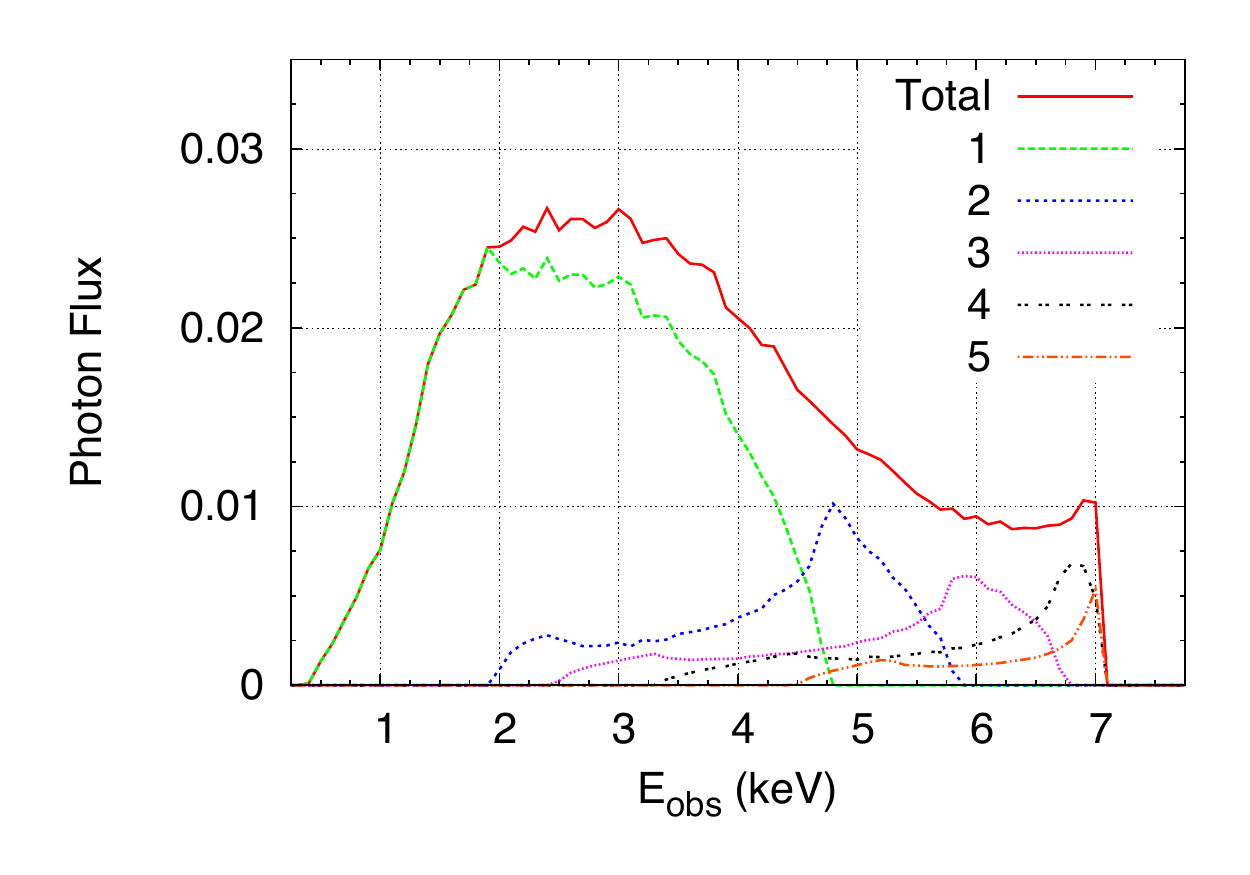}
\end{center}
\vspace{-0.8cm}
\caption{As in Fig.~\ref{fc3} for the configuration~IV KBHSH. \label{fc4}}
\vspace{0.3cm}
\begin{center}
\includegraphics[type=pdf,ext=.pdf,read=.pdf,width=11cm]{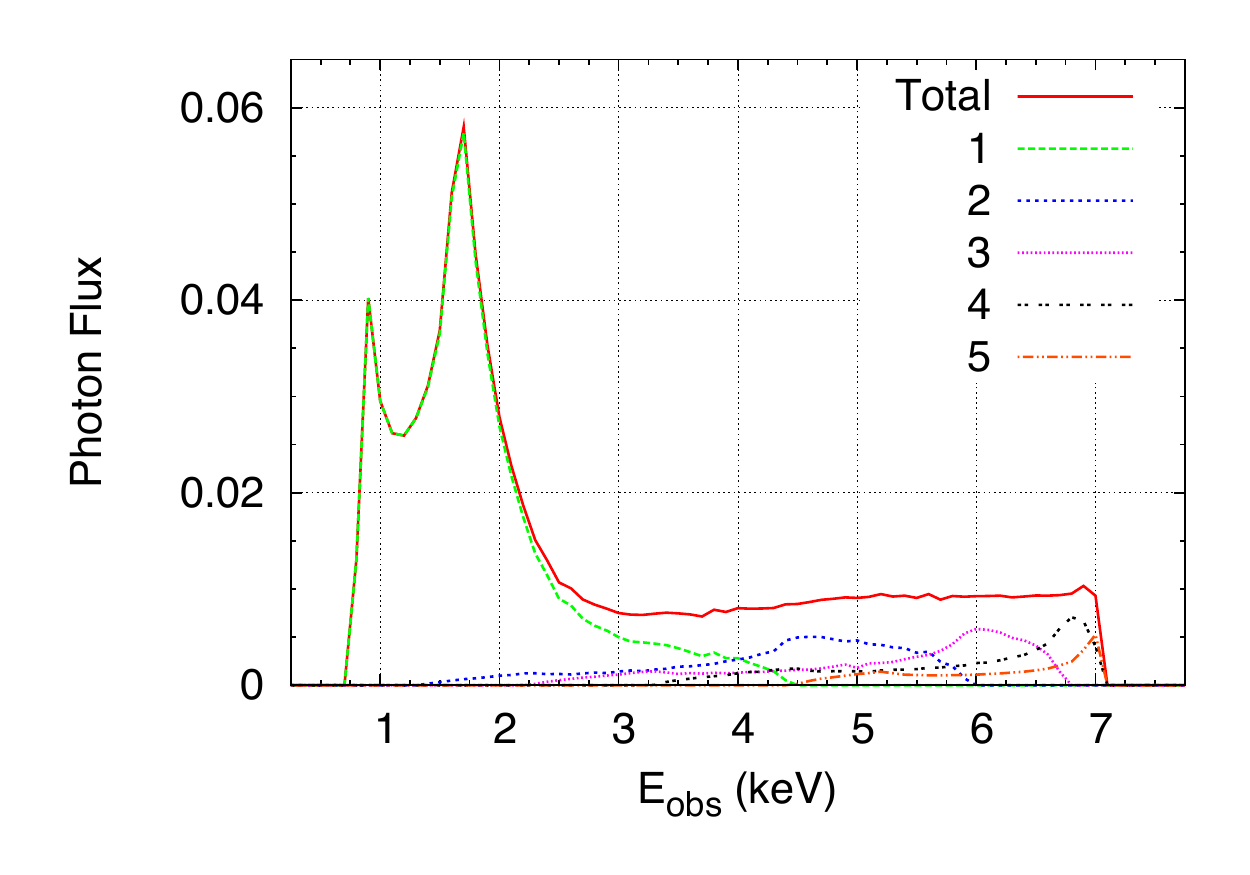}
\end{center}
\vspace{-0.8cm}
\caption{As in Fig.~\ref{fc3} for the configuration~V KBHSH. \label{fc5}}
\end{figure}

The shape of the iron lines in Fig.~\ref{firon} can be better understood if we decompose each iron line into the iron lines produced from different annuli of the accretion disk. This is shown in Figs.~\ref{fc3}, \ref{fc4}, and \ref{fc5}, respectively for the configurations~III, IV, and V. The red solid lines represent the total iron lines already shown in Fig.~\ref{firon}, where the disk has the inner edge at the ISCO radius, i.e. $r_{\rm in} = r_{\rm ISCO}$, and the outer edge at $r_{\rm out} = r_{\rm ISCO} + 25$\footnote{As in previous works on KBHsSH, see e.g. Section~3.3 in Ref.~\cite{Herdeiro:2015gia}, distances are expressed in units of the Compton wave length of the scalar field (rather than Schwarzschild radii). Thus, the radial coordinate is in units of $1/\mu = 1$, where $\mu$ is the scalar field mass (as usual we use $G_{\rm N} = c = \hbar = 1$). Moreover, the ADM masses of the three configurations are reported at the end of Section~\ref{s-2}, as well as the horizon mass that can be much lower when the most of the ADM mass is in the scale field cloud. For example, for the case~III we have $r = r_{\rm ISCO} + M/0.415$.}. The lines indicated by the numbers 1, 2, 3, 4, and 5 corresponds to the iron lines produced from annuli of the accretion disk at increasing radii. Their inner and outer radii are as follows:
\be
\begin{array}{lll}
\text{Annulus~1:} & \quad r_{\rm in} = r_{\rm ISCO} & \quad r_{\rm out} = r_{\rm ISCO} + 1 \\
\text{Annulus~2:} & \quad r_{\rm in} = r_{\rm ISCO} + 1 & \quad r_{\rm out} = r_{\rm ISCO} + 2 \\
\text{Annulus~3:} & \quad r_{\rm in} = r_{\rm ISCO} + 2 & \quad r_{\rm out} = r_{\rm ISCO} + 4 \\
\text{Annulus~4:} & \quad r_{\rm in} = r_{\rm ISCO} + 4 & \quad r_{\rm out} = r_{\rm ISCO} + 10 \\
\text{Annulus~5:} & \quad r_{\rm in} = r_{\rm ISCO} + 10 & \quad r_{\rm out} = r_{\rm ISCO} + 25 \\
\end{array}
\ee
Here and in the next section we employ a ``small'' accretion disk, so our constraints have to be taken with caution (more as a proof of principle indeed).

The iron line of the KBHSH~III is closer to that of a Kerr BH without scalar hair, see Fig.~\ref{fc3}. The contribution from the inner annulus (annulus~1) is relatively moderate: while the emission of radiation at small radii is higher (the local spectrum scales as $1/r^3$), a significant fraction of the photons is swallowed by the central BH. The peak of the photon flux at high energies results from photons emitted at larger radii, where the gravitational redshift is milder.

The iron line of the KBHSH~V is substantially different, see Fig.~\ref{fc5}. The contribution from the inner annulus is very important: the local spectrum still scales as $1/r^3$, but now the photons emitted at small radii can more easily escape to infinity. Physically speaking, this is because the horizon is smaller, as most of the mass is stored in the scalar hair rather than the horizon, and thus the absorption cross section for light is also smaller. The two peaks are produced by the Doppler redshift and blueshift, due to the rotation of the gas. The presence of two peaks at low energy in the KBHSH~V is a feature already found in a large class of exotic compact stars without horizon~\cite{stars} and in different types of traversable wormholes~\cite{menglei,wh}. It is typical of objects without horizon, where only a small fraction of photons emitted near the inner edge of the disk is captured by the central object. The absence of the peak at high energies is because we are plotting a normalized iron line profile, so the number of high energy photons is simply much lower than that of low energy photons. The iron line of the KBHSH~IV in Fig.~\ref{fc4} is something between the two previous cases.

\begin{table}
\begin{center}
\begin{tabular}{|c |c c c |c c c|}
\hline
& & AGN & & & BH Binary & \\
\hline
BH & $\chi^2_{\rm min}$ & $a_*$ & $i$ [deg] & $\chi^2_{\rm min}$ & $a_*$ & $i$ [deg] \\
\hline
Configuration~III & 1.003 & $0.47^{+0.12}_{-0.15}$ & $< 15.9$ & $1.008$ & $> 0.81$ & $< 14.7$ \\
Configuration~IV & 1.034 & $> 0.995$ & $35.1^{+7.8}_{-7.2}$ & $0.983$ & $0.9938^{+0.0032}_{-0.0019}$ & $35.0^{+0.5}_{-2.8}$ \\ 
Configuration~V & 1.216 & $0.9861^{+0.0005}_{-0.0015}$ & $< 6.9$ & $6.08$ & --- & --- \\
\hline
\end{tabular}
\end{center}
\caption{Summary of the fits of the simulations with XIS/Suzaku, respectively for the case of an AGN and a BH binary. The spectra of the configurations III and IV can be fitted with a Kerr model ($\chi^2_{\rm min}$ is close to 1). However, the best fits for the spin parameter $a_*$ and the inclination angle $i$ are wrong (especially for the configuration III); an independent estimate of the inclination angle may potentially reveal the actual nature of the compact object even in the presence of a good fit of its reflection spectrum. In the case of the configuration~V, both the AGN and the binary spectra cannot be fitted with a Kerr model. The folded spectrum and the ratio between the data and the model of the configuration~V are shown in Fig.~\ref{fs1} (AGN) and Fig.~\ref{fs2} (binary).\label{tab1}}
\vspace{0.8cm}
\begin{center}
\begin{tabular}{|c |c c c |c c c|}
\hline
& & AGN & & & BH Binary & \\
\hline
BH & $\chi^2_{\rm min}$ & $a_*$ & $i$ [deg] & $\chi^2_{\rm min}$ & $a_*$ & $i$ [deg] \\
\hline
Configuration~III & 0.993 & $> 0.91$ & $28^{+23}_{-13}$ & $1.793$ & $> 0.990$ & $7.9^{+1.2}_{-0.7}$ \\
Configuration~IV & 1.002 & $> 0.981$ & $30.4^{+17.8}_{-10.0}$ & $11.95$ & --- & --- \\ 
Configuration~V & 2.61 & --- & --- & $3043$ & --- & --- \\
\hline
\end{tabular}
\end{center}
\caption{As in Tab.~\ref{tab1} for the simulations with eXTP. In the case of AGN, we use LFA/eXTP, while in the case of BH binary we use LAD/eXTP. In the BH binary case, the Kerr model can never fit our simulated spectra. The folded spectrum and the ratio between the data and the model of the BH binary cases are shown in Fig.~\ref{fs3} (configuration~III) and Fig.~\ref{fs4} (configuration~IV).\label{tab2}}
\end{table}

\begin{figure}
\begin{center}
\includegraphics[type=pdf,ext=.pdf,read=.pdf,width=16cm]{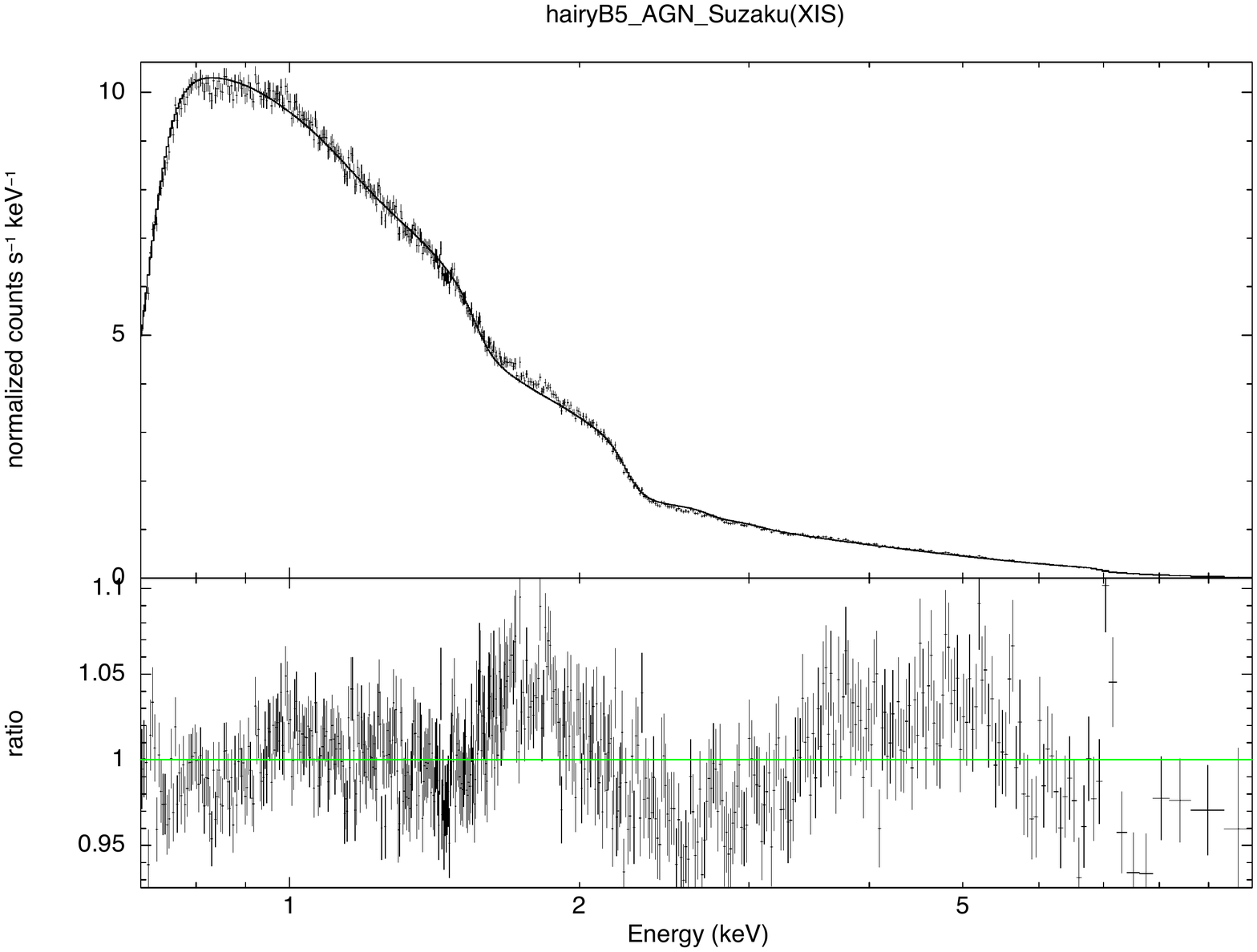}
\end{center}
\vspace{-1.0cm}
\caption{Simulation of the spectrum of a configuration~V KBHSH in an AGN with XIS/Suzaku. Top panel: simulated folded spectrum. Bottom panel: ratio between the simulated data and the Kerr model. Here $\chi^2_{\rm min} = 1.216$. See the text for more details.}
\label{fs1}
\end{figure}

\begin{figure}
\begin{center}
\includegraphics[type=pdf,ext=.pdf,read=.pdf,width=16cm]{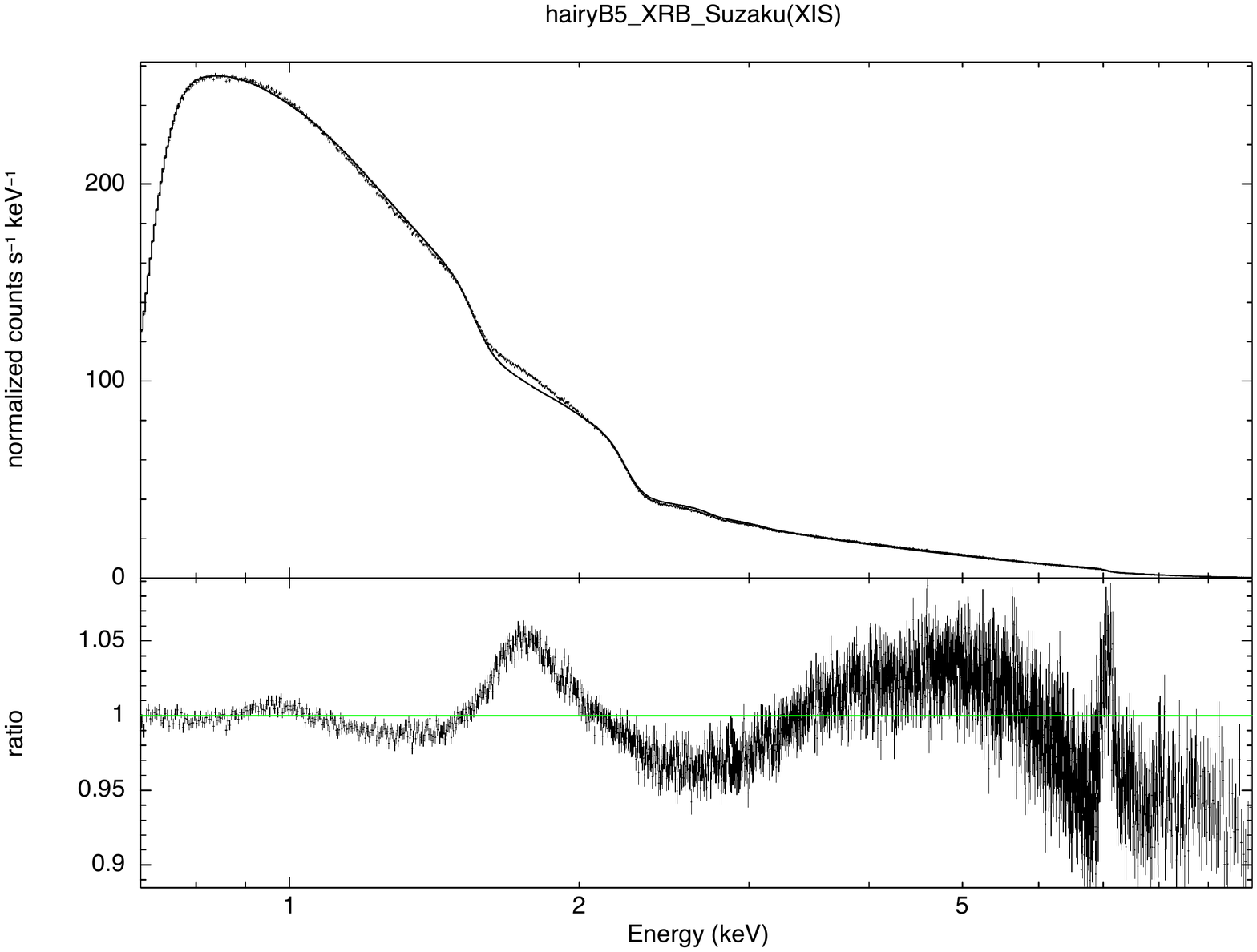}
\end{center}
\vspace{-1.0cm}
\caption{Simulation of the spectrum of a configuration~V KBHSH in an binary with XIS/Suzaku. Top panel: simulated folded spectrum. Bottom panel: ratio between the simulated data and the Kerr model. Here $\chi^2_{\rm min} = 6.08$. See the text for more details.}
\label{fs2}
\end{figure}

\begin{figure}
\begin{center}
\includegraphics[type=pdf,ext=.pdf,read=.pdf,width=16cm]{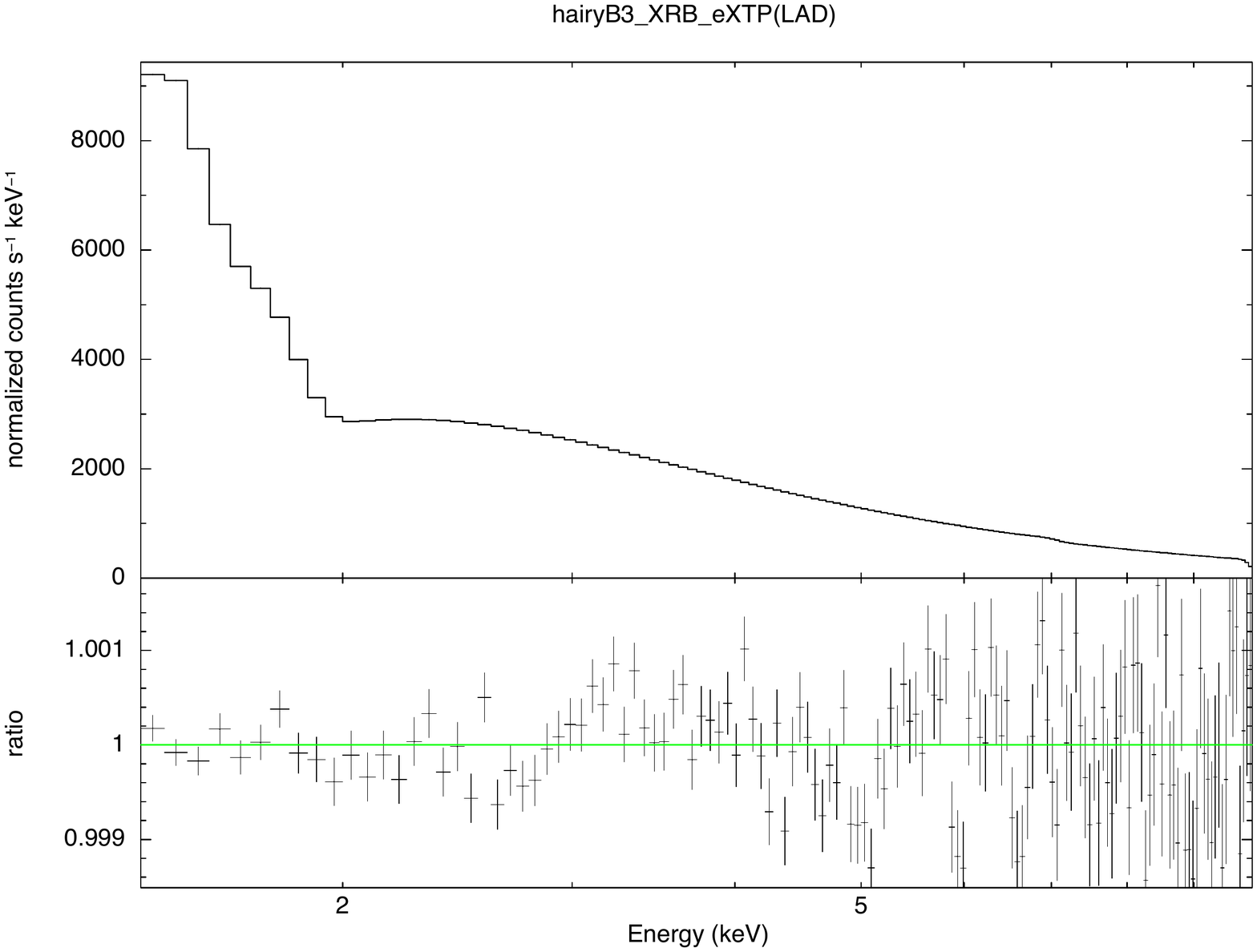}
\end{center}
\vspace{-1.0cm}
\caption{Simulation of the spectrum of a configuration~III KBHSH in a binary with LAD/eXTP. Top panel: simulated folded spectrum. Bottom panel: ratio between the simulated data and the Kerr model. Here $\chi^2_{\rm min} = 1.793$. See the text for more details.}
\label{fs3}
\end{figure}

\begin{figure}
\begin{center}
\includegraphics[type=pdf,ext=.pdf,read=.pdf,width=16cm]{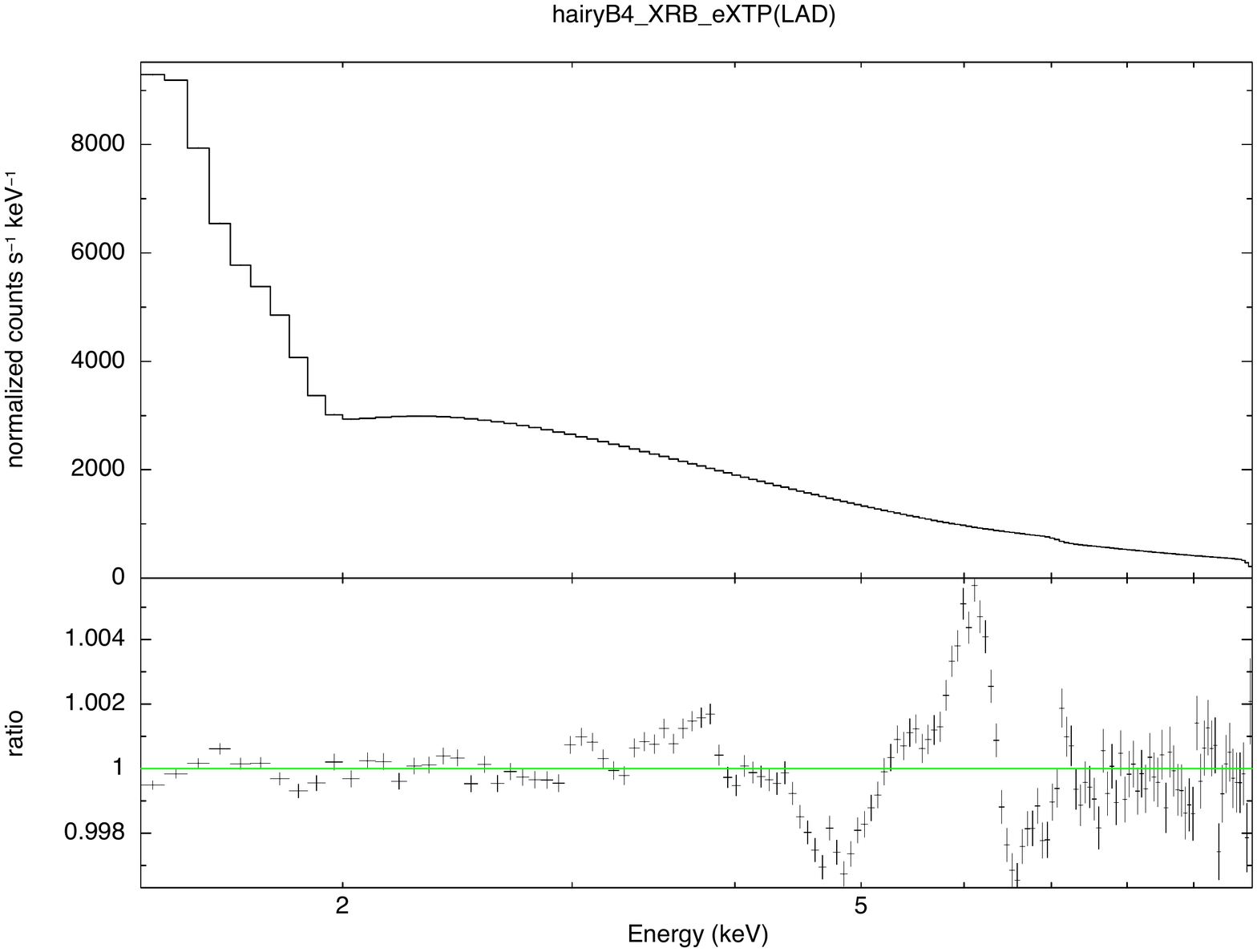}
\end{center}
\vspace{-1.0cm}
\caption{Simulation of the spectrum of a configuration~IV KBHSH in a binary with LAD/eXTP. Top panel: simulated folded spectrum. Bottom panel: ratio between the simulated data and the Kerr model. Here $\chi^2_{\rm min} = 11.95$. See the text for more details.}
\label{fs4}
\end{figure}

\section{Simulations with Suzaku and eXTP \label{s-4}}

In the previous section, we have obtained the iron line profiles expected in the three spacetimes named, respectively, configuration~III, IV, and V. It is clear that, at least in some cases, the iron line profile can be very different from that expected in the reflection spectrum of a Kerr BH. In this section, we want to be more quantitative and check whether present or future observational facilities can constrain the hairiness of the BHs in this model.

Our strategy is as follows. We simulate the X-ray spectrum of a typical AGN and of a typical binary. For the sake of simplicity, we model the spectrum of these sources with a power law with photon index $\Gamma = 2$ (representing the spectrum of a hot corona) and a broad iron line (the reflection spectrum). 
$\Gamma = 2$ is quite a typical value of the photon index for the continuum originated by inverse Compton scattering of thermal photons from the disk off hot electrons in the corona, see, e.g., Ref.~\cite{gamma=2}.
The iron lines used in these simulations are the three iron lines presented in the previous section and obtained in the three KBHsSH metrics assuming a viewing angle $i=45^\circ$ and an emissivity index $\alpha = 3$.

In the case of the AGN, we assume that its energy flux in the 0.7-10~keV range is about $2 \cdot 10^{-10}$~erg/s/cm$^2$. In the case of the X-ray binary, we adopt the value $4 \cdot 10^{-9}$~erg/s/cm$^2$. In both cases, we assume that the iron line has an equivalent width of about 200~eV. These are quite typical parameters for sources suitable to reflection measurements.

We first perform simulations with XIS/Suzaku. We use XSPEC\footnote{http://heasarc.gsfc.nasa.gov/docs/xanadu/xspec/index.html} with the background, ancillary, and the response matrix files of XIS/Suzaku to fake the data. We assume a time exposure of 100~ks, obtaining a photon count of about $1 \cdot 10^6$ in the AGN case, and of about $3 \cdot 10^7$ for the binary. These data are then considered real data and, after rebinning (we impose that the minimum photon count per bin is 20), they are analyzed following the standard procedure. The faked data have been fitted with a power law plus an iron line for a Kerr model. The fitting parameters are: the photon index of the power law $\Gamma$, the normalization of the power law, the rest-frame energy of the iron line, the spin parameter $a_*$, the viewing angle $i$, emissivity index $\alpha_1$, the breaking radius $r_{\rm b}$, and the normalization of the iron line. The emissivity index $\alpha_2$ for $r > r_{\rm b}$ is set to 3, the Newtonian limit for the lamppost geometry.

The summary of the fits of the simulations with XIS/Suzaku for the three hairy BHs is reported in Tab.~\ref{tab1}. The left part of the table is for the AGN, the right one for the binary. The table shows the minimum of $\chi^2$ for any simulation, and the values of the best fit for the spin parameter $a_*$ and the inclination angle $i$. The spectra of the KBHSH of configuration~V cannot be fitted (even in the case of the AGN, $\chi^2_{\rm min} > 1.2$ is not acceptable). The simulated folded spectrum\footnote{The {\it folded spectrum} is the actual spectrum measured by an instrument (in units of counts per spectral bin)
\be
C(h) = \tau \int R(h,E) \, A(E) \, s(E) \, dE \, ,
\ee
where $h$ is the spectral channel, $\tau$ is the exposure time, $R(h,E)$ is the redistribution matrix (essentially the response of the instrument), $A(E)$ is the effective area, and $s(E)$ is the intrinsic spectrum of the source. In general, the redistribution matrix cannot be inverted, and for this reason one usually considers the folded spectrum. The {\it unfolded spectrum} is an attempt to reconstruct the intrinsic spectrum $s(E)$.} and the ratio between the simulated data and the Kerr model are shown in Fig.~\ref{fs1} (AGN) and Fig.~\ref{fs2} (binary). Especially in the case of the binary, it is clear that the iron line profile cannot be fitted with a Kerr model.

We note that our simulations include the power law component from the corona, so we are not considering an ideal and unrealistic observation. We do not take into account any thermal component from the disk, but this is not a bad assumption. In the case of AGN, such a component is in the optical/UV band and it can thus be completely ignored. In the case of binaries, one typically selects observations in which the sources are in the hard state and the temperature of the disk is low, so the thermal component plays a minor rule. We also note that our simulations assume the simplest (and not realistic) case $I_{\rm e} \propto 1/r^3$, but we fit the simulated data with a broken power law. We do not expect that a different profile in the simulations changes our conclusion, but future work should also explore the impact of the uncertainty in the disk corona geometry on the constraints on $q$.

In the case of the spectra of the configurations~III and IV, we obtain a good fit ($\chi^2_{\rm min} \approx 1.0$ and no clear unresolved features) even if the estimate of the inclination angle $i$ may be completely wrong (for configuration~III). The fact that the fits are good means that the faked data can be well fitted with a Kerr model, namely a typical observation of an AGN or a binary today cannot distinguish these objects from Kerr BHs. The fact that the estimate of the inclination angle is wrong in the case of the configuration~III means that there is a degeneracy. In principle, an independent measurement of the inclination angle may solve the degeneracy, but this is not so straightforward and a possible disagreement may also be interpreted with a misalignment between the inner part of the disk with the spin of the BH or the orbital plane of the binary system.

In order to explore the opportunities offered by the next generation of X-ray missions, we consider simulations with eXTP. We assume the same exposure time, namely 100~ks, and we employ LFA/eXTP for the AGN and LAD/eXTP for the binary. The photon counts are, respectively, about $3 \cdot 10^6$ and about $3 \cdot 10^9$. The photon count in the case of the binary is much higher, thanks to the large effective area of LAD. The result of the fits are shown in Tab.~\ref{tab2}. In the AGN case, the spectra of the hairy BHs of the configurations~III and IV can still be fitted with a Kerr model. In the binary case, this is not true and in both cases we get $\chi^2_{\rm min} > 1.2$ and we observe clear unresolved features in the ratio between the data and the Kerr model. The simulated folded spectrum and the ratio between the simulated data and the Kerr model for the binary simulations are shown in Fig.~\ref{fs3} (configuration~III) and Fig.~\ref{fs4} (configuration~IV).

\section{Summary and conclusions \label{s-5}}

Motivated by the uniqueness theorems established for (electro)vacuum, together with some partial theorems in the presence of other matter sources, a consensus that BHs are relatively simple objects,  characterized by a small number of parameters, all of which are measurable at infinity (since they are associated to Gauss laws), settled in. As Wheeler summarized it: \textit{BHs have no hair}. Then, the spacetime geometry around astrophysical BHs should be well approximated by the Kerr metric. Nevertheless, theoretical investigation has shown hairy BHs are possible in the presence of new physics. Thus, caution demands that both the theoretical and phenomenological properties of possible alternative compact objects should be analyzed, before fully committing to the Kerr paradigm, which is especially timely in the current epoch, wherein unprecedented amounts of observational data concerning strong gravity systems is becoming available.

In this paper, we have considered the hairy BHs discovered in Ref.~\cite{hbh} and we have studied possible observational signatures. These BHs appear in 4-dimensional Einstein gravity in the presence of a minimally coupled complex, massive scalar field. Introducing the dimensionless scalar hair $q$ ranging from 0 to 1, these solutions reduce to standard Kerr BHs without scalar hair for $q=0$, while they become boson stars without BH in the opposite limit $q=1$. In the general case with $0 < q < 1$, these solutions describe a central Kerr BH surrounded by a scalar cloud that deforms the geometry.

We have studied the profile of the iron K$\alpha$ line expected in the reflection spectrum of a KBHSH. The iron line method is indeed a powerful tool to probe the spacetime geometry of the near horizon region of an astrophysical BH. As a preliminary study, we have employed a simple model adding an iron line to a power law component, as well as a small accretion disk. We have considered three hairy BH metrics (configuration~III, IV, and V) and we have simulated their observation for the case of an AGN and a BH binary, employing typical parameters for these two kinds of sources. For the simulations, we have adopted XIS/Suzaku as a prototype of a current observation and LFA/eXTP and LAD/eXTP, respectively for the AGN and binary cases, as a prototype of observations with a future X-ray mission. The simulated spectra have been fitted with a Kerr model to check whether it is possible to get a good fit or not. In the presence of a good fit with the Kerr model, we can conclude that a similar hairy BH cannot be distinguished by a standard Kerr BH without scalar hair. In the presence of a bad fit, we can distinguish a Kerr BH with a certain scalar charge from one without scalar charge.

The available X-ray data of astrophysical BHs are ordinarily fitted with Kerr model and there is no tension between data and theoretical predictions. We can thus argue that certain regions in the domain of existence of KBHsSH can be excluded by current data, even if here we have not performed any systematic analysis to infer a bound on the dimensionless scalar hair $q$ and delimit the corresponding region in the domain of existence of KBHsSH. This requires a more thorough scan of such domain, beyond the scope of the present paper, which only establishes a proof of concept that such constraining can be made. Of course, the fact that KBHsSH are only known as numerical solutions, makes this analysis much more challenging, including the computation expected reflection spectrum.

Future X-ray missions, like eXTP, can further tighten the constraints, thanks to their lower background and larger effective area. BH candidates in X-ray binaries may be more suitable for this kind of tests, thanks to the fact they are brighter sources. However, this conclusion should be taken with some caution, because in a real observation the spectrum of a BH binary is usually more complicated to model and systematic effects may prevent a reliable measurement. More work is surely necessary to arrive at clear, quantitative conclusions.


\begin{acknowledgments}
We would like to thank Jiachen Jiang, Andrea Marinucci, Paolo Pani, and James Steiner for useful discussions and suggestions. YN, MZ and CB were supported by the NSFC (grants 11305038 and U1531117) and the Thousand Young Talents Program. CB also acknowledges support from the Alexander von Humboldt Foundation. AC-A acknowledges funding from the Fundaci\'on Universitaria Konrad Lorenz (Project 5INV1161). CH and ER acknowledge funding from the FCT-IF programme. This work was partially supported by  the  H2020-MSCA-RISE-2015 Grant No.  StronGrHEP-690904, and by the CIDMA project UID/MAT/04106/2013.
\end{acknowledgments}


\end{document}